\documentclass[aps,prl,reprint,superscriptaddress]{revtex4-1}
\usepackage{graphicx}
\usepackage{epstopdf}
\providecommand{\e}[1]{\ensuremath{\times 10^{#1}}}

\begin{document}
\title{Extraordinary optical transmission of multimode quantum correlations via localized surface plasmons}
\date{\today}

\author{B.J. Lawrie}
\email{lawriebj@ornl.gov}
\author{P.G. Evans}
\author{R.C. Pooser}
\email{pooserrc@ornl.gov}
\affiliation{Computational Sciences and Engineering Division, Oak Ridge National Laboratory, Oak Ridge, Tennessee 37831, USA}

\begin{abstract}
We demonstrate the transduction of macroscopic quantum correlations by Ag localized surface plasmons (LSPs). Quantum noise reduction, or squeezed light, generated through four-wave-mixing in Rb vapor, is coupled to a Ag nanohole array designed to exhibit LSP-mediated extraordinary-optical transmission (EOT) spectrally coincident with the squeezed light source at 795 nm. This first demonstration of the coupling of quantum light into LSPs conserves spatially dependent quantum information, allowing for parallel quantum protocols in on-chip sub-wavelength quantum information processing.\end{abstract}

\maketitle


The growth of applications in plasmonics ranging from nano-imaging \cite{Kawata2009} to sub-wavelength photonic circuits \cite{Ozbay2006} has motivated the growing interest in quantum plasmonics \cite{Altewischer2002,Chen2011,Martino2012,Lee2012,Huck2009,Jacob2011}.  Recent reports of nanoscale beamsplitters, phase shifters, and crossover splitters \cite{Yurke2010} based on localized surface plasmons (LSPs) supported on metal nanoparticle networks provide the basis of a quantum information nano-circuit, provided that the coherent transduction of quantum information into LSPs is feasible.  In addition, sensors utilizing quantum noise reduction (otherwise known as ``squeezed light'') demonstrate sensitivity below the photon shot noise limit, a particularly useful feature in low-light applications \cite{Caves1981,Goda2008}.  Recent literature has demonstrated a sensitivity at the standard quantum limit (SQL) of 4\e{-9} refractive index units for plasmonic biosensors based on extraordinary optical transmission (EOT) \cite{Ebbesen1998} of classical light sources through sub-wavelength hole arrays \cite{Yang2009, Leebeeck2007}.  The ability to coherently couple squeezed light through sub-wavelength hole arrays will allow for significant improvement in the sensitivity of EOT based biosensors, particularly for sensors utilizing photosensitive ligands and demonstrating low-light intensity EOT saturation \cite{Zayats2003}.  In this letter, we provide the first demonstration of the transduction of a squeezed state through an EOT medium, and the first demonstration of the transduction of quantum information by LSPs.

\begin{figure}[!hb]
\includegraphics[width=\columnwidth]{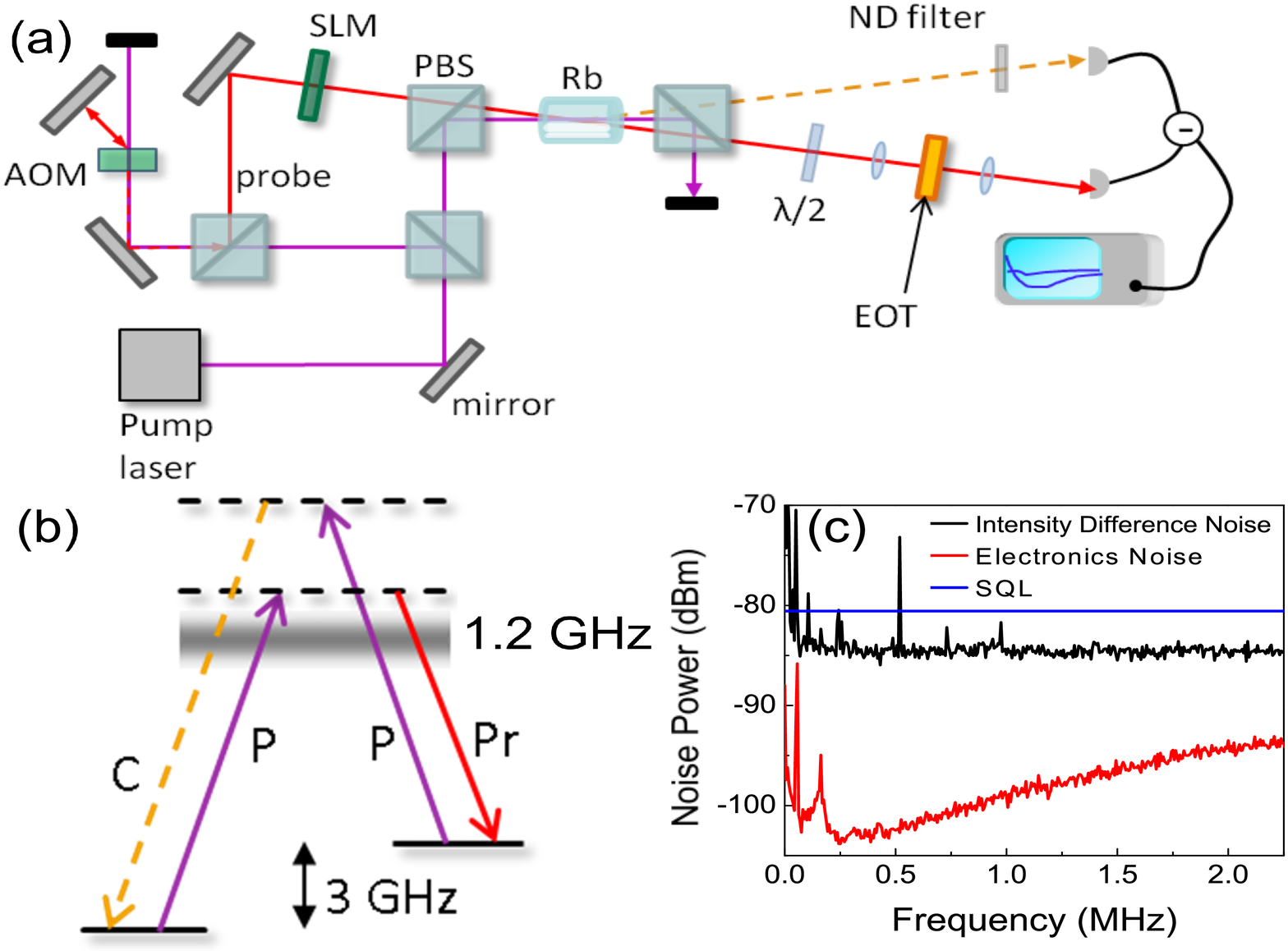}
\caption[fig1]{\label{fig:fig1}(Color online) (a) A schematic of the 4WM experiment with abbreviations SLM: spatial light modulator, PBS: polarizing beam splitter, AOM: acousto-optic modulator, and ND: neutral density.   (b) The energy level diagram for the D1 line in Rb at 795 nm, showing two pump (P) photons absorbed by the vapor and two subsequent probe (Pr) and conjugate (C) photons re-emitted with quantum correlations. (c) A typical squeezing spectrum showing a maximum of 4 dB of squeezing at frequencies above 1 MHz.  This squeezing spectrum was acquired with 10\% attenuation on the probe and conjugate corresponding to the transmission through an indium tin oxide coated glass substrate.}
\end{figure}

Huck \emph{et al.} recently demonstrated the transduction of squeezed light into surface plasmon polaritons (SPPs) supported in Au waveguides \cite{Huck2009}, and Altewischer \emph{et al.} demonstrated the transduction of entangled photons by SPPs through SPP-mediated extraordinary optical transmission \cite{Altewischer2002}.  However, because SPP excitation is highly dependent on the incident photon wavevector, quantum decoherence is observed when polarization entangled photon pairs are focused to undergo SPP-mediated EOT \cite{Altewischer2002,Moreno2004,Altewischer2005}.  LSPs are fundamentally different excitations from SPPs: characterized by discrete resonances dependent on the size, shape and dielectric function of the nanostructured material, LSPs can decay efficiently through photon emission independent of the incident wavevector \cite{Zayats2003}.  While the wavevector dependence of SPP excitation and decay has resulted in difficulties in quantum SPP imaging \cite{Altewischer2002}, the transduction of quantum information into LSPs would suffer from no such issue \cite{Zayats2003}, and as a result, quantum LSPs can function as a workbench for quantum nano-imaging and quantum information in plasmonic systems.


In order to demonstrate the transduction of squeezed light into LSPs, a two-mode squeezed state is generated by four wave mixing (4WM) in ${}^{85}$Rb vapor as illustrated schematically in Fig.~1a. The nonlinear four wave mixing interaction is based on a double lambda system between the hyperfine ground states and the excited states in  ${}^{85}$Rb at the D1 line (795 nm) \cite{McCormick2007OL,McCormick2008,Boyer2008}.  The ${}^{85}$Rb vapor absorbs two photons from the pump beam (denoted by ``P'' in Fig.~1b), which builds coherence between the two ground state levels. The presence of a probe beam (``pr'' in Fig.~1b) stimulates the vapor to re-emit photons into the probe and a third beam called the conjugate (``C'' in Fig.~1b). The process is coherent, and the probe and conjugate photons are emitted simultaneously, resulting in quantum correlations that are observed in the form of intensity difference squeezing 4.5 dB below the SQL.  The SQL was determined as a function of incident power via measurements of the noise present in a coherent light source at powers matching the sum of the probe and conjugate powers.

The absence of a cavity in atomic vapor 4WM configurations results in a squeezed light source that is insensitive to environmental noise and capable of producing multi-spatial-mode squeezed light \cite{Boyer2008,Marino2009}.  In order to take advantage of this, a spatial light modulator (SLM) is used prior to the Rb vapor cell in order to provide control over which spatial modes are supported on the probe beam.  A double-pass acousto-optic modulator is used to offset the probe frequency from the pump by 3.045 GHz, close to the frequency spacing of the ground state hyperfine splitting, and the pump frequency is detuned 1.2 GHz to the blue of the  ${}^{85}$Rb ${\mathrm F}=2\rightarrow {\mathrm F}^\prime$  transition. As currently configured, this source provides 4.5 dB of quantum noise reduction.  In order to examine the degree to which squeezing was transmitted by LSPs, the probe was focused with a 5 cm lens on the nanohole array prior to collection by the detector.  A variable neutral density filter was used in the conjugate beampath to balance the classical intensity noise after attenuation of the probe by the EOT heterostructure.  A typical squeezing spectrum as a function of sideband frequency for light coupled through the nanohole array substrate is shown in Fig.~1(c). 

Recently, other authors have demonstrated LSP-mediated EOT of classical light in nanostructured hole arrays of triangles \cite{Rodrigo2010} and crosses \cite{Lin2011}. The transmission spectra of similar triangular nanohole geometries were modeled by finite difference time domain (FDTD) simulations in order to determine the nanohole array dimensions required for maximum transmission at 795 nm.  Fig.~2a shows the simulated transmission spectra, as a function of incident polarization state, for nano-triangular holes in an 80-nm thick Ag film on fused silica. The triangles have bases 200 nm in length, legs 288 nm in length, and a pitch of 400 nm. The electric field profile measured 10 nm below the hole array at the peak near-IR transmission wavelength for polarizations of $0^{\circ}$ and $60^{\circ}$ (where $\vec{E}$ is parallel to the triangle base for $0^{\circ}$ polarization) illustrates the localized nature of the excitation responsible for the EOT near 800 nm, consistent with the reports in the literature for EOT in similar heterostructures \cite{Rodrigo2010}.

\begin{figure}
\includegraphics[width=\columnwidth]{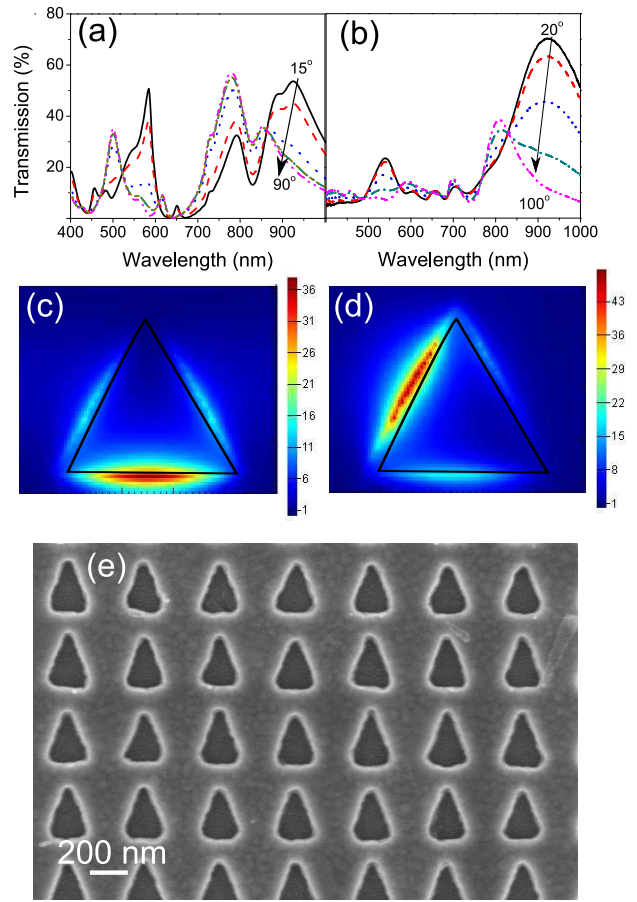}
\caption[fig2]{\label{fig:fig2}(Color online) Simulated (a) and experimental (b) transmission spectrum for triangle hole arrays with bases of 230 nm, legs 300 nm in size, and a pitch of 400 nm in 80 nm Ag thin film on indium tin oxide coated fused silica for polarization varying from $15^{\circ}$ to $90^{\circ}$ in increments of $15^{\circ}$ (simulated) and from $20^{\circ}$ to $100^{\circ}$ in increments of $20^{\circ}$ (experimental).  The 200 nm x 200 nm electric field profile 10 nm below the Ag film is shown in (c) and (d) for polarizations of $0^{\circ}$ and $60^{\circ}$. (e) An SEM image of a representative portion of a  200 $\mu$m $\times$ 200 $\mu$m triangle hole array fabricated based on the optimal geometry found by FDTD simulations.}
\end{figure}

\begin{figure}
\includegraphics[width=\columnwidth]{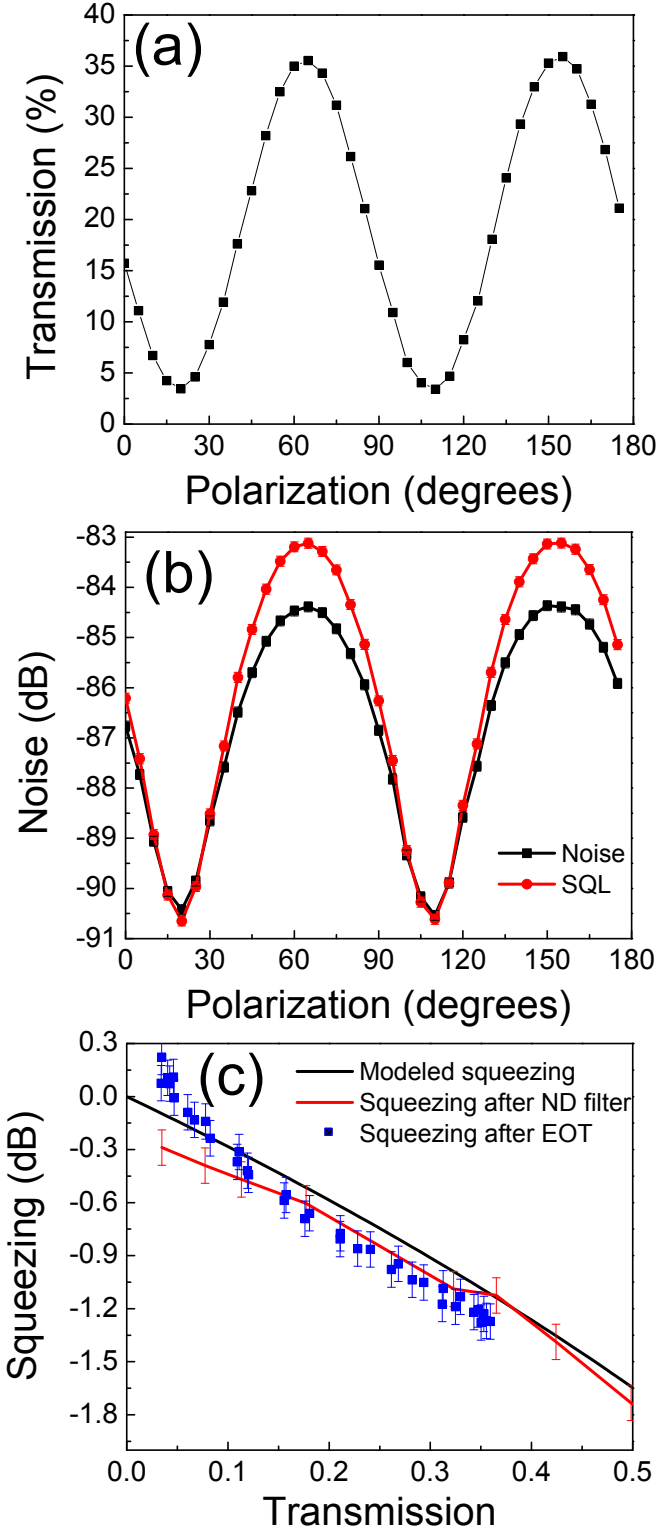}
\caption[fig3]{\label{fig:data}(Color online) (a) Single color transmission for hole array for polarization of $0-175^{\circ}$, and (b) the relative intensity noise and standard quantum limit for the rebalanced probe and conjugate after probe has passed through the EOT medium. (c) plots the measured squeezing as a function of transmission through the hole array and through a variable neutral density filter.  The noise based on the model shown in eq. 3 is shown in black.  The error bars in (a) and some error bars in (b) are smaller than the symbols.}
\end{figure}

Heterostructures comprising a square lattice of isosceles triangle holes, having base $b=200$ nm, legs $l=287$ nm, and pitch $p=400$ nm, were fabricated using an electron-beam lithography process on indium-tin-oxide coated borosilicate glass.  A bilayer process was developed using Poly(methyl methacrylate) 495k A4 and 950k A2 resists in order to provide for the necessary undercut required for successful lift-off after the electron beam evaporation of an 80 nm Ag film. Fig.~2e shows a scanning electron microscope image of the triangle hole arrays at 100kx magnification.  The transmission spectra of a representative nanohole array was measured with a standard confocal transmission microscope and the spectra were normalized to include the 90\% transmission through the indium tin oxide coated substrate.  Aside from some discrepancies in amplitude and evidence of minor inhomogenous broadening, the spectral positions of the simulated transmission spectra in Fig.~2a correspond qualitatively with the experimental transmission spectra shown in Fig.~2b. 

The transmission for the heterostructure illustrated in Fig.~2e was reacquired as a single color spectrum as a function of incident polarization, as shown in Fig.~3(a), in order to verify the transmission of the film at the wavelength of our quantum light source. The peak transmission of 35.9\% transmission at a polarization of $65^{\circ}$ is consistent with the FDTD simulations, while the minimum transmission of 3\% at a polarization of $20^{\circ}$ is consistent with the existence of a strong (1,1) mode Wood-Rayleigh anomaly \cite{Wood1902}. In contrast, the transmission through an 80 nm Ag thin film without a hole array is less than 1\% for wavelengths between 300 nm and 1000 nm. 

Having established the polarization dependence of EOT for these nanostructured hole arrays, Fig.~3(b) illustrates the the average intensity difference noise measured between 1.25 MHz and 2.25 MHz and the shot noise at each polarization.  The noise shown in Fig.~3 is the average of 4\e4 data points within the 1 MHz bandwidth, so that, in concert with the systematic error due largely to laser drift, the total error in the measured squeezing is ~0.1 dB.  The maximum squeezing observed is 1.28 dB at polarizations of $60^{\circ}$ and $150^{\circ}$, coincident with the 36\% transmission.

The effects of ohmic losses in SPP waveguides \cite{Ballester2009} and the effects of scattering in metal nanoparticle arrays \cite{Lee2012} on quantum information have been previously treated theoretically via effective beam splitters.  We explored this effective beamsplitter model for our geometry as follows.  A variable neutral density filter was used to simultaneously attenuate the probe and conjugate in order to provide a direct comparison between the squeezing transfer by LSP-mediated EOT and by a partially reflective filter.  In addition, the squeezing was treated as the result of a single amplifier with gain G followed by a beamsplitter with transmission of $\eta$ so that the probe and conjugate field operators $a_{1out}$ and $a_{2out}$ are given by:

\begin{equation}
a_{1out}=\sqrt{\eta G}a_{1in}+\sqrt{\eta (G-1)}a_{2in}^\dagger+\sqrt{1-\eta}a_{vd1}
 \label{eq:a1
}
\end{equation}
\begin{equation}
a_{2out}=\sqrt{\eta G}a_{2in}+\sqrt{\eta (G-1)}a_{1in}^\dagger+\sqrt{1-\eta}a_{vd2}
 \label{eq:a2
}
\end{equation}
where $a_{vd1,2}$ correspond to the input vaccum fields associated with the second beam splitter port.
The difference noise is given by $\Delta (a_{1} ^\dagger \cdot a_{1}-a_{2} ^\dagger \cdot a_{2}){^2} = \langle (a_{1} ^\dagger \cdot a_{1}-a_{2} ^\dagger \cdot a_{2}){^2} \rangle-\langle a_{1} ^\dagger \cdot a_{1}-a_{2} ^\dagger \cdot a_{2}\rangle^2$, which yields relative intensity noise of:

\begin{equation}
{\Delta (a_{1} ^\dagger \cdot a_{1}-a_{2} ^\dagger \cdot a_{2})}{^2}=1-\eta+\eta/G
 \label{eq:squeezing
}
\end{equation}

after normalizing by the shot noise for each output mode.  Fig.~3(c) plots the squeezing as a function of transmission through a variable attenuator for a gain of G=4 and for transmission that is normalized against the 30\% attenuation resulting from near-resonance absorption in the Rb vapor cell and the 94\% efficient photodetector.  

While the data acquired with a ND  filter in both probe and conjugate beam-paths coincide well with the model based on attenuation of quantum correlations by an effective beam splitter, the EOT-mediated squeezing drops significantly for transmission of less than 11\%.  Indeed, at a polarization of $20^{\circ}$, the EOT data demonstrates apparent anti-squeezing of 0.2 dB, a significant deviation from the  0.29 dB of squeezing observed when the ND filter was used to attenuate the probe and conjugate equally.  In contrast to the results seen for SPP-mediated EOT, no change to the gaussian beam profile was observed after the probe was coupled through the hole array.  Indeed, when the image of a cross was placed on the probe beam by a spatial light modulator, the image of the cross propagated through the hole array with no observable degradation of image quality - as seen in Fig. 4 - and no loss of squeezing compared with the gaussian beam profile.  This experimentally demonstrates the capability for LSP-mediated EOT to transmit multiple spatial modes and images effectively while conserving quantum information. 

\begin{figure}
\includegraphics[width=\columnwidth]{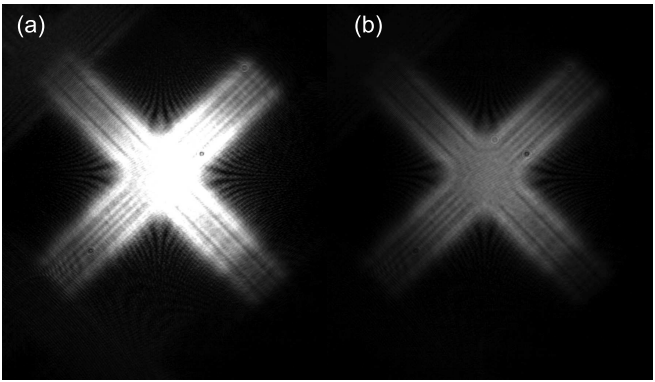}
\caption[fig4]{\label{fig:fig4}Beam profile of the probe beam taken under identical camera settings both without (a) and with (b) transduction by the EOT medium.}
\end{figure}

In this paper, we have demonstrated the effective transduction of multi-spatial-mode intensity squeezed light by LSPs in nanostructured hole arrays.  For applications varying from sub-shot-noise EOT based biosensors to quantum plasmonic circuitry, the ability to effectively excite continuous quantum variable LSPs is a topic of great significance.  In conjunction with the current state of the art squeezing of 8-12 dB \cite{McCormick2008,Eberle2010}, improved transmission comparable to the 75\% transmission observed at 950 nm in Fig. 2(b) will allow for EOT sensors demonstrating dramatic sensitivity improvement beyond the shot noise limit.  Because they support multiple spatial modes, the LSP-mediated EOT heterostructures discussed here should continue to be a workbench for the analysis of quantum nanoscale imaging in a way that plasmonic waveguides and SPP-mediated EOT heterostructures could not. The deviation of the squeezing from effective beam splitter models previously used for theoretical predictions of quantum information attenuation in heterostructures supporting both LSPs and SPPs will require further study. While a full theoretical treatment of this deviation is beyond the scope of this paper, we anticipate that a forthcoming study detailing the spatial mode dependence of squeezing transduction by LSPs will clarify this issue.

\begin{acknowledgments}
This work was performed at Oak Ridge National Laboratory, operated by UT-Battelle for the U.S. Department of energy under contract no. DE-AC05-00OR22725.  A portion of this research was conducted at the Center for Nanophase Materials Sciences, which is sponsored at Oak Ridge National Laboratory by the Scientific User Facilities Division, Office of Basic Energy Sciences, U.S. Department of Energy.  B.J.L was supported by a fellowship from the IC postdoctoral research program. R.C.P. acknowledges partial support from the Wigner fellowship.
\end{acknowledgments}


\begin{thebibliography}{26}%
\makeatletter
\providecommand \@ifxundefined [1]{%
 \@ifx{#1\undefined}
}%
\providecommand \@ifnum [1]{%
 \ifnum #1\expandafter \@firstoftwo
 \else \expandafter \@secondoftwo
 \fi
}%
\providecommand \@ifx [1]{%
 \ifx #1\expandafter \@firstoftwo
 \else \expandafter \@secondoftwo
 \fi
}%
\providecommand \natexlab [1]{#1}%
\providecommand \enquote  [1]{``#1''}%
\providecommand \bibnamefont  [1]{#1}%
\providecommand \bibfnamefont [1]{#1}%
\providecommand \citenamefont [1]{#1}%
\providecommand \href@noop [0]{\@secondoftwo}%
\providecommand \href [0]{\begingroup \@sanitize@url \@href}%
\providecommand \@href[1]{\@@startlink{#1}\@@href}%
\providecommand \@@href[1]{\endgroup#1\@@endlink}%
\providecommand \@sanitize@url [0]{\catcode `\\12\catcode `\$12\catcode
  `\&12\catcode `\#12\catcode `\^12\catcode `\_12\catcode `\%12\relax}%
\providecommand \@@startlink[1]{}%
\providecommand \@@endlink[0]{}%
\providecommand \url  [0]{\begingroup\@sanitize@url \@url }%
\providecommand \@url [1]{\endgroup\@href {#1}{\urlprefix }}%
\providecommand \urlprefix  [0]{URL }%
\providecommand \Eprint [0]{\href }%
\providecommand \doibase [0]{http://dx.doi.org/}%
\providecommand \selectlanguage [0]{\@gobble}%
\providecommand \bibinfo  [0]{\@secondoftwo}%
\providecommand \bibfield  [0]{\@secondoftwo}%
\providecommand \translation [1]{[#1]}%
\providecommand \BibitemOpen [0]{}%
\providecommand \bibitemStop [0]{}%
\providecommand \bibitemNoStop [0]{.\EOS\space}%
\providecommand \EOS [0]{\spacefactor3000\relax}%
\providecommand \BibitemShut  [1]{\csname bibitem#1\endcsname}%
\let\auto@bib@innerbib\@empty
\bibitem [{\citenamefont {{Kawata}}\ \emph {et~al.}(2009)\citenamefont
  {{Kawata}}, \citenamefont {{Inouye}},\ and\ \citenamefont
  {{Verma}}}]{Kawata2009}%
  \BibitemOpen
  \bibfield  {author} {\bibinfo {author} {\bibfnamefont {S.}~\bibnamefont
  {{Kawata}}}, \bibinfo {author} {\bibfnamefont {Y.}~\bibnamefont {{Inouye}}},
  \ and\ \bibinfo {author} {\bibfnamefont {P.}~\bibnamefont {{Verma}}},\ }\href
  {\doibase 10.1038/nphoton.2009.111} {\bibfield  {journal} {\bibinfo
  {journal} {Nature Photonics}\ }\textbf {\bibinfo {volume} {3}},\ \bibinfo
  {pages} {388} (\bibinfo {year} {2009})}\BibitemShut {NoStop}%
\bibitem [{\citenamefont {Ozbay}(2006)}]{Ozbay2006}%
  \BibitemOpen
  \bibfield  {author} {\bibinfo {author} {\bibfnamefont {E.}~\bibnamefont
  {Ozbay}},\ }\href {\doibase 10.1126/science.1114849} {\bibfield  {journal}
  {\bibinfo  {journal} {Science}\ }\textbf {\bibinfo {volume} {311}},\ \bibinfo
  {pages} {189} (\bibinfo {year} {2006})}\BibitemShut {NoStop}%
\bibitem [{\citenamefont {Altewischer}\ \emph {et~al.}(2002)\citenamefont
  {Altewischer}, \citenamefont {van Exter},\ and\ \citenamefont
  {Woerdman}}]{Altewischer2002}%
  \BibitemOpen
  \bibfield  {author} {\bibinfo {author} {\bibfnamefont {E.}~\bibnamefont
  {Altewischer}}, \bibinfo {author} {\bibfnamefont {M.}~\bibnamefont {van
  Exter}}, \ and\ \bibinfo {author} {\bibfnamefont {J.}~\bibnamefont
  {Woerdman}},\ }\href@noop {} {\bibfield  {journal} {\bibinfo  {journal}
  {Nature}\ }\textbf {\bibinfo {volume} {418}},\ \bibinfo {pages} {304}
  (\bibinfo {year} {2002})}\BibitemShut {NoStop}%
\bibitem [{\citenamefont {Chen}\ \emph {et~al.}(2011)\citenamefont {Chen},
  \citenamefont {Lambert}, \citenamefont {Chou}, \citenamefont {Chen},\ and\
  \citenamefont {Nori}}]{Chen2011}%
  \BibitemOpen
  \bibfield  {author} {\bibinfo {author} {\bibfnamefont {G.-Y.}\ \bibnamefont
  {Chen}}, \bibinfo {author} {\bibfnamefont {N.}~\bibnamefont {Lambert}},
  \bibinfo {author} {\bibfnamefont {C.-H.}\ \bibnamefont {Chou}}, \bibinfo
  {author} {\bibfnamefont {Y.-N.}\ \bibnamefont {Chen}}, \ and\ \bibinfo
  {author} {\bibfnamefont {F.}~\bibnamefont {Nori}},\ }\href {\doibase
  10.1103/PhysRevB.84.045310} {\bibfield  {journal} {\bibinfo  {journal} {Phys.
  Rev. B}\ }\textbf {\bibinfo {volume} {84}},\ \bibinfo {pages} {045310}
  (\bibinfo {year} {2011})}\BibitemShut {NoStop}%
\bibitem [{\citenamefont {Di~Martino}\ \emph {et~al.}(2012)\citenamefont
  {Di~Martino}, \citenamefont {Sonnefraud}, \citenamefont {K\'{e}na-Cohen},
  \citenamefont {Tame}, \citenamefont {\"{O}zdemir}, \citenamefont {Kim},\ and\
  \citenamefont {Maier}}]{Martino2012}%
  \BibitemOpen
  \bibfield  {author} {\bibinfo {author} {\bibfnamefont {G.}~\bibnamefont
  {Di~Martino}}, \bibinfo {author} {\bibfnamefont {Y.}~\bibnamefont
  {Sonnefraud}}, \bibinfo {author} {\bibfnamefont {S.}~\bibnamefont
  {K\'{e}na-Cohen}}, \bibinfo {author} {\bibfnamefont {M.}~\bibnamefont
  {Tame}}, \bibinfo {author} {\bibfnamefont {S.~K.}\ \bibnamefont
  {\"{O}zdemir}}, \bibinfo {author} {\bibfnamefont {M.~S.}\ \bibnamefont
  {Kim}}, \ and\ \bibinfo {author} {\bibfnamefont {S.~A.}\ \bibnamefont
  {Maier}},\ }\href {\doibase 10.1021/nl300671w} {\bibfield  {journal}
  {\bibinfo  {journal} {Nano Letters}\ }\textbf {\bibinfo {volume} {12}},\
  \bibinfo {pages} {2504} (\bibinfo {year} {2012})}\BibitemShut {NoStop}%
\bibitem [{\citenamefont {Lee}\ \emph {et~al.}(2012)\citenamefont {Lee},
  \citenamefont {Tame}, \citenamefont {Lim},\ and\ \citenamefont
  {Lee}}]{Lee2012}%
  \BibitemOpen
  \bibfield  {author} {\bibinfo {author} {\bibfnamefont {C.}~\bibnamefont
  {Lee}}, \bibinfo {author} {\bibfnamefont {M.}~\bibnamefont {Tame}}, \bibinfo
  {author} {\bibfnamefont {J.}~\bibnamefont {Lim}}, \ and\ \bibinfo {author}
  {\bibfnamefont {J.}~\bibnamefont {Lee}},\ }\href {\doibase
  10.1103/PhysRevA.85.063823} {\bibfield  {journal} {\bibinfo  {journal} {Phys.
  Rev. A}\ }\textbf {\bibinfo {volume} {85}},\ \bibinfo {pages} {063823}
  (\bibinfo {year} {2012})}\BibitemShut {NoStop}%
\bibitem [{\citenamefont {Huck}\ \emph {et~al.}(2009)\citenamefont {Huck},
  \citenamefont {Smolka}, \citenamefont {Lodahl}, \citenamefont {S\o{}rensen},
  \citenamefont {Boltasseva}, \citenamefont {Janousek},\ and\ \citenamefont
  {Andersen}}]{Huck2009}%
  \BibitemOpen
  \bibfield  {author} {\bibinfo {author} {\bibfnamefont {A.}~\bibnamefont
  {Huck}}, \bibinfo {author} {\bibfnamefont {S.}~\bibnamefont {Smolka}},
  \bibinfo {author} {\bibfnamefont {P.}~\bibnamefont {Lodahl}}, \bibinfo
  {author} {\bibfnamefont {A.~S.}\ \bibnamefont {S\o{}rensen}}, \bibinfo
  {author} {\bibfnamefont {A.}~\bibnamefont {Boltasseva}}, \bibinfo {author}
  {\bibfnamefont {J.}~\bibnamefont {Janousek}}, \ and\ \bibinfo {author}
  {\bibfnamefont {U.~L.}\ \bibnamefont {Andersen}},\ }\href {\doibase
  10.1103/PhysRevLett.102.246802} {\bibfield  {journal} {\bibinfo  {journal}
  {Phys. Rev. Lett.}\ }\textbf {\bibinfo {volume} {102}},\ \bibinfo {pages}
  {246802} (\bibinfo {year} {2009})}\BibitemShut {NoStop}%
\bibitem [{\citenamefont {Jacob}\ and\ \citenamefont
  {Shalaev}(2011)}]{Jacob2011}%
  \BibitemOpen
  \bibfield  {author} {\bibinfo {author} {\bibfnamefont {Z.}~\bibnamefont
  {Jacob}}\ and\ \bibinfo {author} {\bibfnamefont {V.}~\bibnamefont
  {Shalaev}},\ }\href@noop {} {\bibfield  {journal} {\bibinfo  {journal}
  {Science}\ }\textbf {\bibinfo {volume} {334}},\ \bibinfo {pages} {463}
  (\bibinfo {year} {2011})}\BibitemShut {NoStop}%
\bibitem [{\citenamefont {Yurke}\ and\ \citenamefont
  {Kuang}(2010)}]{Yurke2010}%
  \BibitemOpen
  \bibfield  {author} {\bibinfo {author} {\bibfnamefont {B.}~\bibnamefont
  {Yurke}}\ and\ \bibinfo {author} {\bibfnamefont {W.}~\bibnamefont {Kuang}},\
  }\href {\doibase 10.1103/PhysRevA.81.033814} {\bibfield  {journal} {\bibinfo
  {journal} {Phys. Rev. A}\ }\textbf {\bibinfo {volume} {81}},\ \bibinfo
  {pages} {033814} (\bibinfo {year} {2010})}\BibitemShut {NoStop}%
\bibitem [{\citenamefont {Caves}(1981)}]{Caves1981}%
  \BibitemOpen
  \bibfield  {author} {\bibinfo {author} {\bibfnamefont {C.~M.}\ \bibnamefont
  {Caves}},\ }\href {\doibase 10.1103/PhysRevD.23.1693} {\bibfield  {journal}
  {\bibinfo  {journal} {Phys. Rev. D}\ }\textbf {\bibinfo {volume} {23}},\
  \bibinfo {pages} {1693} (\bibinfo {year} {1981})}\BibitemShut {NoStop}%
\bibitem [{\citenamefont {Goda}\ \emph {et~al.}(2008)\citenamefont {Goda},
  \citenamefont {Miyakawa}, \citenamefont {Mikhailov}, \citenamefont {Saraf},
  \citenamefont {Adhikari}, \citenamefont {McKenzie}, \citenamefont {Ward},
  \citenamefont {Vass}, \citenamefont {Weinstein},\ and\ \citenamefont
  {Mavalvala}}]{Goda2008}%
  \BibitemOpen
  \bibfield  {author} {\bibinfo {author} {\bibfnamefont {K.}~\bibnamefont
  {Goda}}, \bibinfo {author} {\bibfnamefont {O.}~\bibnamefont {Miyakawa}},
  \bibinfo {author} {\bibfnamefont {E.}~\bibnamefont {Mikhailov}}, \bibinfo
  {author} {\bibfnamefont {S.}~\bibnamefont {Saraf}}, \bibinfo {author}
  {\bibfnamefont {R.}~\bibnamefont {Adhikari}}, \bibinfo {author}
  {\bibfnamefont {K.}~\bibnamefont {McKenzie}}, \bibinfo {author}
  {\bibfnamefont {R.}~\bibnamefont {Ward}}, \bibinfo {author} {\bibfnamefont
  {S.}~\bibnamefont {Vass}}, \bibinfo {author} {\bibfnamefont {A.}~\bibnamefont
  {Weinstein}}, \ and\ \bibinfo {author} {\bibfnamefont {N.}~\bibnamefont
  {Mavalvala}},\ }\href@noop {} {\bibfield  {journal} {\bibinfo  {journal}
  {Nature Physics}\ }\textbf {\bibinfo {volume} {4}},\ \bibinfo {pages} {472}
  (\bibinfo {year} {2008})}\BibitemShut {NoStop}%
\bibitem [{\citenamefont {Ebbesen}\ \emph {et~al.}(1998)\citenamefont
  {Ebbesen}, \citenamefont {Lezec}, \citenamefont {Ghaemi}, \citenamefont
  {Thio},\ and\ \citenamefont {Wolff}}]{Ebbesen1998}%
  \BibitemOpen
  \bibfield  {author} {\bibinfo {author} {\bibfnamefont {T.}~\bibnamefont
  {Ebbesen}}, \bibinfo {author} {\bibfnamefont {H.}~\bibnamefont {Lezec}},
  \bibinfo {author} {\bibfnamefont {H.}~\bibnamefont {Ghaemi}}, \bibinfo
  {author} {\bibfnamefont {T.}~\bibnamefont {Thio}}, \ and\ \bibinfo {author}
  {\bibfnamefont {P.}~\bibnamefont {Wolff}},\ }\href@noop {} {\bibfield
  {journal} {\bibinfo  {journal} {Nature}\ }\textbf {\bibinfo {volume} {391}},\
  \bibinfo {pages} {667} (\bibinfo {year} {1998})}\BibitemShut {NoStop}%
\bibitem [{\citenamefont {Yang}\ and\ \citenamefont {Ho}(2009)}]{Yang2009}%
  \BibitemOpen
  \bibfield  {author} {\bibinfo {author} {\bibfnamefont {T.}~\bibnamefont
  {Yang}}\ and\ \bibinfo {author} {\bibfnamefont {H.~P.}\ \bibnamefont {Ho}},\
  }\href {\doibase 10.1364/OE.17.011205} {\bibfield  {journal} {\bibinfo
  {journal} {Opt. Express}\ }\textbf {\bibinfo {volume} {17}},\ \bibinfo
  {pages} {11205} (\bibinfo {year} {2009})}\BibitemShut {NoStop}%
\bibitem [{\citenamefont {De~Leebeeck}\ \emph {et~al.}(2007)\citenamefont
  {De~Leebeeck}, \citenamefont {Kumar}, \citenamefont {de~Lange}, \citenamefont
  {Sinton}, \citenamefont {Gordon},\ and\ \citenamefont
  {Brolo}}]{Leebeeck2007}%
  \BibitemOpen
  \bibfield  {author} {\bibinfo {author} {\bibfnamefont {A.}~\bibnamefont
  {De~Leebeeck}}, \bibinfo {author} {\bibfnamefont {L.~K.~S.}\ \bibnamefont
  {Kumar}}, \bibinfo {author} {\bibfnamefont {V.}~\bibnamefont {de~Lange}},
  \bibinfo {author} {\bibfnamefont {D.}~\bibnamefont {Sinton}}, \bibinfo
  {author} {\bibfnamefont {R.}~\bibnamefont {Gordon}}, \ and\ \bibinfo {author}
  {\bibfnamefont {A.~G.}\ \bibnamefont {Brolo}},\ }\href {\doibase
  10.1021/ac070001a} {\bibfield  {journal} {\bibinfo  {journal} {Analytical
  Chemistry}\ }\textbf {\bibinfo {volume} {79}},\ \bibinfo {pages} {4094}
  (\bibinfo {year} {2007})}\BibitemShut {NoStop}%
\bibitem [{\citenamefont {Zayats}\ and\ \citenamefont
  {Smolyaninov}(2003)}]{Zayats2003}%
  \BibitemOpen
  \bibfield  {author} {\bibinfo {author} {\bibfnamefont {A.~V.}\ \bibnamefont
  {Zayats}}\ and\ \bibinfo {author} {\bibfnamefont {I.~I.}\ \bibnamefont
  {Smolyaninov}},\ }\href {http://stacks.iop.org/1464-4258/5/i=4/a=353}
  {\bibfield  {journal} {\bibinfo  {journal} {Journal of Optics A: Pure and
  Applied Optics}\ }\textbf {\bibinfo {volume} {5}},\ \bibinfo {pages} {S16}
  (\bibinfo {year} {2003})}\BibitemShut {NoStop}%
\bibitem [{\citenamefont {Moreno}\ \emph {et~al.}(2004)\citenamefont {Moreno},
  \citenamefont {Garc{\'\i}a-Vidal}, \citenamefont {Erni}, \citenamefont
  {Cirac},\ and\ \citenamefont {Mart{\'\i}n-Moreno}}]{Moreno2004}%
  \BibitemOpen
  \bibfield  {author} {\bibinfo {author} {\bibfnamefont {E.}~\bibnamefont
  {Moreno}}, \bibinfo {author} {\bibfnamefont {F.}~\bibnamefont
  {Garc{\'\i}a-Vidal}}, \bibinfo {author} {\bibfnamefont {D.}~\bibnamefont
  {Erni}}, \bibinfo {author} {\bibfnamefont {J.}~\bibnamefont {Cirac}}, \ and\
  \bibinfo {author} {\bibfnamefont {L.}~\bibnamefont {Mart{\'\i}n-Moreno}},\
  }\href@noop {} {\bibfield  {journal} {\bibinfo  {journal} {Physical Review
  Letters}\ }\textbf {\bibinfo {volume} {92}},\ \bibinfo {pages} {236801}
  (\bibinfo {year} {2004})}\BibitemShut {NoStop}%
\bibitem [{\citenamefont {Altewischer}\ \emph {et~al.}(2005)\citenamefont
  {Altewischer}, \citenamefont {Oei}, \citenamefont {van Exter},\ and\
  \citenamefont {Woerdman}}]{Altewischer2005}%
  \BibitemOpen
  \bibfield  {author} {\bibinfo {author} {\bibfnamefont {E.}~\bibnamefont
  {Altewischer}}, \bibinfo {author} {\bibfnamefont {Y.~C.}\ \bibnamefont
  {Oei}}, \bibinfo {author} {\bibfnamefont {M.~P.}\ \bibnamefont {van Exter}},
  \ and\ \bibinfo {author} {\bibfnamefont {J.~P.}\ \bibnamefont {Woerdman}},\
  }\href {\doibase 10.1103/PhysRevA.72.013817} {\bibfield  {journal} {\bibinfo
  {journal} {Phys. Rev. A}\ }\textbf {\bibinfo {volume} {72}},\ \bibinfo
  {pages} {013817} (\bibinfo {year} {2005})}\BibitemShut {NoStop}%
\bibitem [{\citenamefont {McCormick}\ \emph {et~al.}(2007)\citenamefont
  {McCormick}, \citenamefont {Boyer}, \citenamefont {Arimondo},\ and\
  \citenamefont {Lett}}]{McCormick2007OL}%
  \BibitemOpen
  \bibfield  {author} {\bibinfo {author} {\bibfnamefont {C.~F.}\ \bibnamefont
  {McCormick}}, \bibinfo {author} {\bibfnamefont {V.}~\bibnamefont {Boyer}},
  \bibinfo {author} {\bibfnamefont {E.}~\bibnamefont {Arimondo}}, \ and\
  \bibinfo {author} {\bibfnamefont {P.~D.}\ \bibnamefont {Lett}},\ }\href
  {\doibase 10.1364/OL.32.000178} {\bibfield  {journal} {\bibinfo  {journal}
  {Opt. Lett.}\ }\textbf {\bibinfo {volume} {32}},\ \bibinfo {pages} {178}
  (\bibinfo {year} {2007})}\BibitemShut {NoStop}%
\bibitem [{\citenamefont {McCormick}\ \emph {et~al.}(2008)\citenamefont
  {McCormick}, \citenamefont {Marino}, \citenamefont {Boyer},\ and\
  \citenamefont {Lett}}]{McCormick2008}%
  \BibitemOpen
  \bibfield  {author} {\bibinfo {author} {\bibfnamefont {C.~F.}\ \bibnamefont
  {McCormick}}, \bibinfo {author} {\bibfnamefont {A.~M.}\ \bibnamefont
  {Marino}}, \bibinfo {author} {\bibfnamefont {V.}~\bibnamefont {Boyer}}, \
  and\ \bibinfo {author} {\bibfnamefont {P.~D.}\ \bibnamefont {Lett}},\ }\href
  {\doibase 10.1103/PhysRevA.78.043816} {\bibfield  {journal} {\bibinfo
  {journal} {Phys. Rev. A}\ }\textbf {\bibinfo {volume} {78}},\ \bibinfo
  {pages} {043816} (\bibinfo {year} {2008})}\BibitemShut {NoStop}%
\bibitem [{\citenamefont {Boyer}\ \emph {et~al.}(2008)\citenamefont {Boyer},
  \citenamefont {Marino}, \citenamefont {Pooser},\ and\ \citenamefont
  {Lett}}]{Boyer2008}%
  \BibitemOpen
  \bibfield  {author} {\bibinfo {author} {\bibfnamefont {V.}~\bibnamefont
  {Boyer}}, \bibinfo {author} {\bibfnamefont {A.~M.}\ \bibnamefont {Marino}},
  \bibinfo {author} {\bibfnamefont {R.~C.}\ \bibnamefont {Pooser}}, \ and\
  \bibinfo {author} {\bibfnamefont {P.~D.}\ \bibnamefont {Lett}},\ }\href
  {\doibase 10.1126/science.1158275} {\bibfield  {journal} {\bibinfo  {journal}
  {Science}\ }\textbf {\bibinfo {volume} {321}},\ \bibinfo {pages} {544}
  (\bibinfo {year} {2008})}\BibitemShut {NoStop}%
\bibitem [{\citenamefont {Marino}\ \emph {et~al.}(2009)\citenamefont {Marino},
  \citenamefont {Pooser}, \citenamefont {Boyer},\ and\ \citenamefont
  {Lett}}]{Marino2009}%
  \BibitemOpen
  \bibfield  {author} {\bibinfo {author} {\bibfnamefont {A.}~\bibnamefont
  {Marino}}, \bibinfo {author} {\bibfnamefont {R.}~\bibnamefont {Pooser}},
  \bibinfo {author} {\bibfnamefont {V.}~\bibnamefont {Boyer}}, \ and\ \bibinfo
  {author} {\bibfnamefont {P.}~\bibnamefont {Lett}},\ }\href@noop {} {\bibfield
   {journal} {\bibinfo  {journal} {Nature}\ }\textbf {\bibinfo {volume}
  {457}},\ \bibinfo {pages} {859} (\bibinfo {year} {2009})}\BibitemShut
  {NoStop}%
\bibitem [{\citenamefont {Rodrigo}\ \emph {et~al.}(2010)\citenamefont
  {Rodrigo}, \citenamefont {Mahboub}, \citenamefont {Degiron}, \citenamefont
  {Genet}, \citenamefont {Garc\'{i}a-Vidal}, \citenamefont
  {Mart\'{i}n-Moreno},\ and\ \citenamefont {Ebbesen}}]{Rodrigo2010}%
  \BibitemOpen
  \bibfield  {author} {\bibinfo {author} {\bibfnamefont {S.~G.}\ \bibnamefont
  {Rodrigo}}, \bibinfo {author} {\bibfnamefont {O.}~\bibnamefont {Mahboub}},
  \bibinfo {author} {\bibfnamefont {A.}~\bibnamefont {Degiron}}, \bibinfo
  {author} {\bibfnamefont {C.}~\bibnamefont {Genet}}, \bibinfo {author}
  {\bibfnamefont {F.~J.}\ \bibnamefont {Garc\'{i}a-Vidal}}, \bibinfo {author}
  {\bibfnamefont {L.}~\bibnamefont {Mart\'{i}n-Moreno}}, \ and\ \bibinfo
  {author} {\bibfnamefont {T.~W.}\ \bibnamefont {Ebbesen}},\ }\href {\doibase
  10.1364/OE.18.023691} {\bibfield  {journal} {\bibinfo  {journal} {Opt.
  Express}\ }\textbf {\bibinfo {volume} {18}},\ \bibinfo {pages} {23691}
  (\bibinfo {year} {2010})}\BibitemShut {NoStop}%
\bibitem [{\citenamefont {Lin}\ and\ \citenamefont {Roberts}(2011)}]{Lin2011}%
  \BibitemOpen
  \bibfield  {author} {\bibinfo {author} {\bibfnamefont {L.}~\bibnamefont
  {Lin}}\ and\ \bibinfo {author} {\bibfnamefont {A.}~\bibnamefont {Roberts}},\
  }\href {\doibase 10.1364/OE.19.002626} {\bibfield  {journal} {\bibinfo
  {journal} {Opt. Express}\ }\textbf {\bibinfo {volume} {19}},\ \bibinfo
  {pages} {2626} (\bibinfo {year} {2011})}\BibitemShut {NoStop}%
\bibitem [{\citenamefont {Wood}(1902)}]{Wood1902}%
  \BibitemOpen
  \bibfield  {author} {\bibinfo {author} {\bibfnamefont {R.~W.}\ \bibnamefont
  {Wood}},\ }\href {http://stacks.iop.org/1478-7814/18/i=1/a=325} {\bibfield
  {journal} {\bibinfo  {journal} {Proceedings of the Physical Society of
  London}\ }\textbf {\bibinfo {volume} {18}},\ \bibinfo {pages} {269} (\bibinfo
  {year} {1902})}\BibitemShut {NoStop}%
\bibitem [{\citenamefont {Ballester}\ \emph {et~al.}(2009)\citenamefont
  {Ballester}, \citenamefont {Tame}, \citenamefont {Lee}, \citenamefont {Lee},\
  and\ \citenamefont {Kim}}]{Ballester2009}%
  \BibitemOpen
  \bibfield  {author} {\bibinfo {author} {\bibfnamefont {D.}~\bibnamefont
  {Ballester}}, \bibinfo {author} {\bibfnamefont {M.~S.}\ \bibnamefont {Tame}},
  \bibinfo {author} {\bibfnamefont {C.}~\bibnamefont {Lee}}, \bibinfo {author}
  {\bibfnamefont {J.}~\bibnamefont {Lee}}, \ and\ \bibinfo {author}
  {\bibfnamefont {M.~S.}\ \bibnamefont {Kim}},\ }\href {\doibase
  10.1103/PhysRevA.79.053845} {\bibfield  {journal} {\bibinfo  {journal} {Phys.
  Rev. A}\ }\textbf {\bibinfo {volume} {79}},\ \bibinfo {pages} {053845}
  (\bibinfo {year} {2009})}\BibitemShut {NoStop}%
\bibitem [{\citenamefont {Eberle}\ \emph {et~al.}(2010)\citenamefont {Eberle},
  \citenamefont {Steinlechner}, \citenamefont {Bauchrowitz}, \citenamefont
  {H\"andchen}, \citenamefont {Vahlbruch}, \citenamefont {Mehmet},
  \citenamefont {M\"uller-Ebhardt},\ and\ \citenamefont
  {Schnabel}}]{Eberle2010}%
  \BibitemOpen
  \bibfield  {author} {\bibinfo {author} {\bibfnamefont {T.}~\bibnamefont
  {Eberle}}, \bibinfo {author} {\bibfnamefont {S.}~\bibnamefont
  {Steinlechner}}, \bibinfo {author} {\bibfnamefont {J.}~\bibnamefont
  {Bauchrowitz}}, \bibinfo {author} {\bibfnamefont {V.}~\bibnamefont
  {H\"andchen}}, \bibinfo {author} {\bibfnamefont {H.}~\bibnamefont
  {Vahlbruch}}, \bibinfo {author} {\bibfnamefont {M.}~\bibnamefont {Mehmet}},
  \bibinfo {author} {\bibfnamefont {H.}~\bibnamefont {M\"uller-Ebhardt}}, \
  and\ \bibinfo {author} {\bibfnamefont {R.}~\bibnamefont {Schnabel}},\ }\href
  {\doibase 10.1103/PhysRevLett.104.251102} {\bibfield  {journal} {\bibinfo
  {journal} {Phys. Rev. Lett.}\ }\textbf {\bibinfo {volume} {104}},\ \bibinfo
  {pages} {251102} (\bibinfo {year} {2010})}\BibitemShut {NoStop}%
\end{thebibliography}
\end{document}